\documentclass[grl]{agutex}

\usepackage{graphicx} 

\authorrunninghead{ZENITANI ET AL.}

\titlerunninghead{DISSIPATION REGION IN RECONNECTION}

\authoraddr{Seiji Zenitani,
National Astronomical Observatory of Japan, 2-21-1 Osawa,
Mitaka, Tokyo 181-8588, Japan. (seiji.zenitani@nao.ac.jp)}
\authoraddr{Iku Shinohara,
Institute of Space and Astronautical Science, Japan Aerospace Exploration Agency, 3-1-1 Yoshinodai, Chuo, Sagamihara, Kanagawa 252-5210, Japan (iku@stp.isas.jaxa.jp)}
\authoraddr{Tsugunobu Nagai,
Tokyo Institute of Technology, Tokyo 152-8551, Japan (nagai@geo.titech.ac.jp)}

\begin{document}

\title{Evidence for the dissipation region in magnetotail reconnection}


\authors{Seiji Zenitani, \altaffilmark{1}
Iku Shinohara, \altaffilmark{2}
and Tsugunobu Nagai \altaffilmark{3}}

\altaffiltext{1}
{National Astronomical Observatory of Japan}
\altaffiltext{2}
{Institute of Space and Astronautical Science, Japan Aerospace Exploration Agency}
\altaffiltext{3}
{Tokyo Institute of Technology}

\begin{abstract}
Signatures of the dissipation region of collisionless magnetic reconnection
are investigated by the Geotail spacecraft for the 15 May 2003 event.
The energy dissipation
in the rest frame of the electron's bulk flow is considered
in an approximate form $D^*_e$,
which is validated by a particle-in-cell simulation. 
The dissipation measure is directly evaluated
from the {plasma moments},
the electric field, and the magnetic field.
Using $D^*_e$,
a compact dissipation region is successfully
detected in the vicinity of the possible X-point in Geotail data.
The dissipation rate is $45 {\rm pWm^{-3}}$.
The length of the dissipation region is estimated to
$1$--$2 d_i^{\rm loc}$ (local ion inertial length).
The Lorentz work $W$,
the work rate by Lorentz force to plasmas,
is also introduced. 
It is positive over the reconnection region and
it has a peak around the pileup region away from the X-point.
These new measures $D^*_e$ and $W$ provide
useful information
to understand the reconnection structure. 
\end{abstract}

%
%

%

\begin{article}

%
%

\section{Introduction}

Magnetic reconnection is one of
the most important processes in many plasma systems.
It drives various explosive events
such as solar flares, substorms in the Earth's magnetosphere, and
disruptions in laboratory devices.
It is well known that
the reconnection process is crucially influenced by
the dissipation physics near the reconnection point (X-point). 
The structure of the dissipation region as well as the accommodated physics
is of critical importance for the understanding of the reconnection mechanism.
Significant efforts have been made on this subject
by theories, numerical simulations, and in situ satellite observations.

In a collisionless kinetic plasma,
typically at a distance of the ion inertial length from the X-point,
ions decouple from magnetic field lines while electrons remain magnetized.
As a consequence, Hall physics plays a role inside the ion-decoupling region. 
Signatures of Hall physics
such as quadruple magnetic field,
bipolar normal electric field,
and the relevant current loops \citep{sonnerup79}
were confirmed by {in situ} observations 
in the terrestrial magnetosphere \citep{nagai01,oieroset01}.

Deep inside the ion-decoupling region,
there is a thin layer where electrons depart from field lines. 
A compact dissipation region is located
in a close vicinity of the X-point.
In addition, recent simulation works suggest that
narrow electron jets extend from there,
stretching the nonideal layer in the outflow directions \citep{kari07,shay07}.
These signatures were recently confirmed by satellite observations. 
\citet{nagai11} found
a compact region with an intense cross-tail current
and
neighboring electron jets at a super-Alfv\'{e}nic speed
in a magnetotail reconnection event.

Until recently it was not clear
how to identify the dissipation region
inside the electron nonideal layer.
Taking the energy transfer into account,
\citet{zeni11c,zeni11d} have proposed a general measure of the dissipation region.
Using particle-in-cell (PIC) simulations,
they {distinguished} the dissipation region from the outer electron jet,
which turned out to be an oblique projection of
a non-dissipative current sheet \citep{hesse08,klimas12}.

In this Letter,
we show {in situ} observational evidence for the dissipation region
in a magnetotail reconnection event.

\section{Observation and Analysis}

\begin{figure}[tbp]
\begin{center}
\includegraphics[width={\columnwidth},clip]{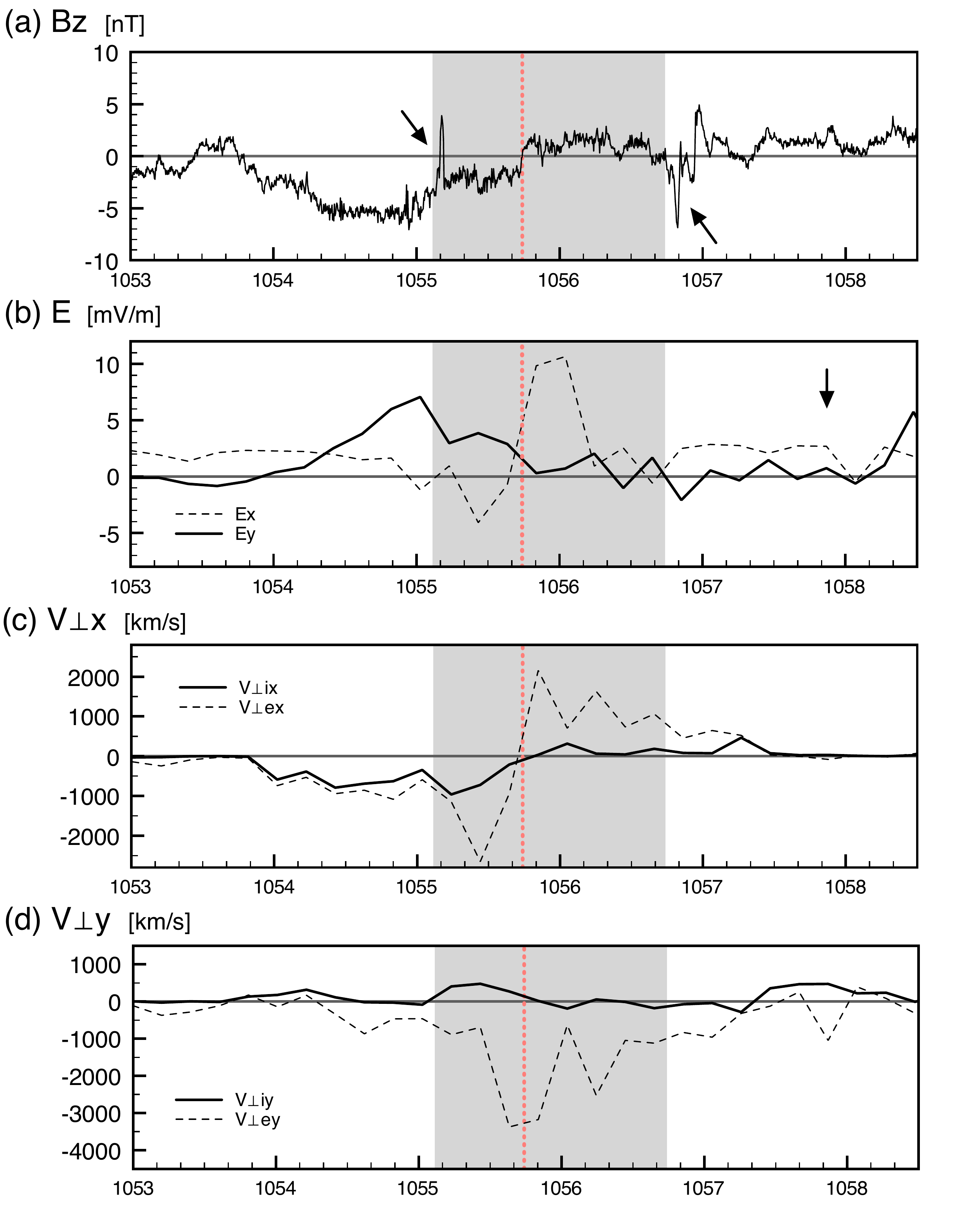}
\caption{(Color online)
Geotail observation {from} 1053:00 UT to 1058:30 UT on 15 May 2003.
(a) Normal magnetic field $B_z$ [nT],
(b) electric field $E_x$, $E_y$ [mV/m], and
(c,d) plasma {perpendicular} velocities [km/s] in the $X_{\rm SC}$ and $Y_{\rm SC}$ directions.
A potential X-point (red dotted line) and
a characteristic interval from 1055:07 to 1056:44 UT (shadow) are overplotted.
}
\end{center}
\label{fig:geotail}
\end{figure}

On 15 May 2003, Geotail detected a reconnection event in the premidnight tail.
This event was extensively analyzed by \citet{nagai11} and so
we only repeat important features here.
Figure \ref{fig:geotail} shows key properties
from 1053:00 UT to 1058:30 UT on 15 May 2003.
All data are presented in the spacecraft (SC) coordinates,
because the electric field data are available only in the SC coordinates.
During the period of our interest,
the SC coordinates are virtually the same as the GSE and GSM coordinates.
The rotation angle to the GSE coordinates is less than $2.4^\circ$ and
the angle to the GSM is less than $5^\circ$. 
The satellite position was
$(X_{\rm GSM}, Y_{\rm GSM}, Z_{\rm GSM}) = (-27.8, +3.3, +3.5 R_{\rm E})$ at 1056 UT.

Figure \ref{fig:geotail}a shows magnetic field $B_z$ at 16 Hz,
measured from the magnetic field experiment (MGF) \citep{kokubun94}.
It suddenly turned from southward ($B_z<0$) to northward ($B_z>0$) at 1055:44 UT,
as indicated by the red dotted lines.
Judging from this and many other signatures \citep{nagai11},
Geotail encountered a potential X-point (hereafter X-point) at this time.

Other quantities are shown at 12 s time resolution.
Figure \ref{fig:geotail}b {shows} electric field data,
obtained from the electric field detector (EFD) \citep{tsuruda94}. 
Technically, they were measured every 3 s and then averaged over 12 s. 
We note that 
the average electric field during 1057:45 -- 1057:58 UT (the arrow in Fig. \ref{fig:geotail}b)
{is} represented by one during 1057:45 -- 1057:52 UT,
because EFD {had}
violent, high-frequency noises of $\Delta E \sim 50 {\rm mVm^{-1}}$
during 1057:52 -- 1057:58 UT.
One can see that $E_x$ has a bipolar signature across the X-point.
This is a {well-known} signature of kinetic reconnection (e.g., \citet{hoshino05}).

Figures \ref{fig:geotail}c and \ref{fig:geotail}d show
the ion and electron bulk velocities.
Plasma {moments} are calculated from distribution functions
that are measured by the low-energy particle experiment (LEP) \citep{mukai94}.
Due to the low time-resolution of 12 s,
these {moments} contain field-aligned flows along the separatrices
{from} 1055:30 UT to 1057:00 UT.  
To rule out the field-aligned Hall currents,
transverse velocities are presented (${\bf v}_{{\perp}i}, {\bf v}_{{\perp}e}$).
The ${v}_{{\perp}ey}$ is distinctly fast
near the X-point (Fig. \ref{fig:geotail}d).
As seen in Figure \ref{fig:geotail}c,
both ${v}_{{\perp}ix}$ and ${v}_{{\perp}ex}$ change
their signs across the X-point.
Note that ${v}_{{\perp}ex}$ overshoots
${v}_{{\perp}ix}$ on both two sides of the X-point.
The outward electron velocity is even faster than
the typical upstream Alfv\'{e}n velocity.
The upstream magnetic field ($10$ nT) and
density ($0.01~{\rm cm}^{-3}$) give $2200$ km/s.
Such bi-directional, super-Alfv\'{e}nic electron jets are
consistent with previous simulations \citep{kari07,shay07}.

The shadows indicate a characteristic interval from 1055:07 to 1056:44 UT,
surrounded by sudden spikes in $B_z$ (Fig. \ref{fig:geotail}a).
This interval features
super-Alfv\'{e}nic electron jets (Fig. \ref{fig:geotail}c),
an intense duskward electron flow (Fig. \ref{fig:geotail}d), and
a low plasma density (not shown).
\citet{nagai11} referred to this interval
as the ``ion-electron decoupling time.''

\begin{figure}[tbp]
\begin{center}
\includegraphics[width={\columnwidth},clip]{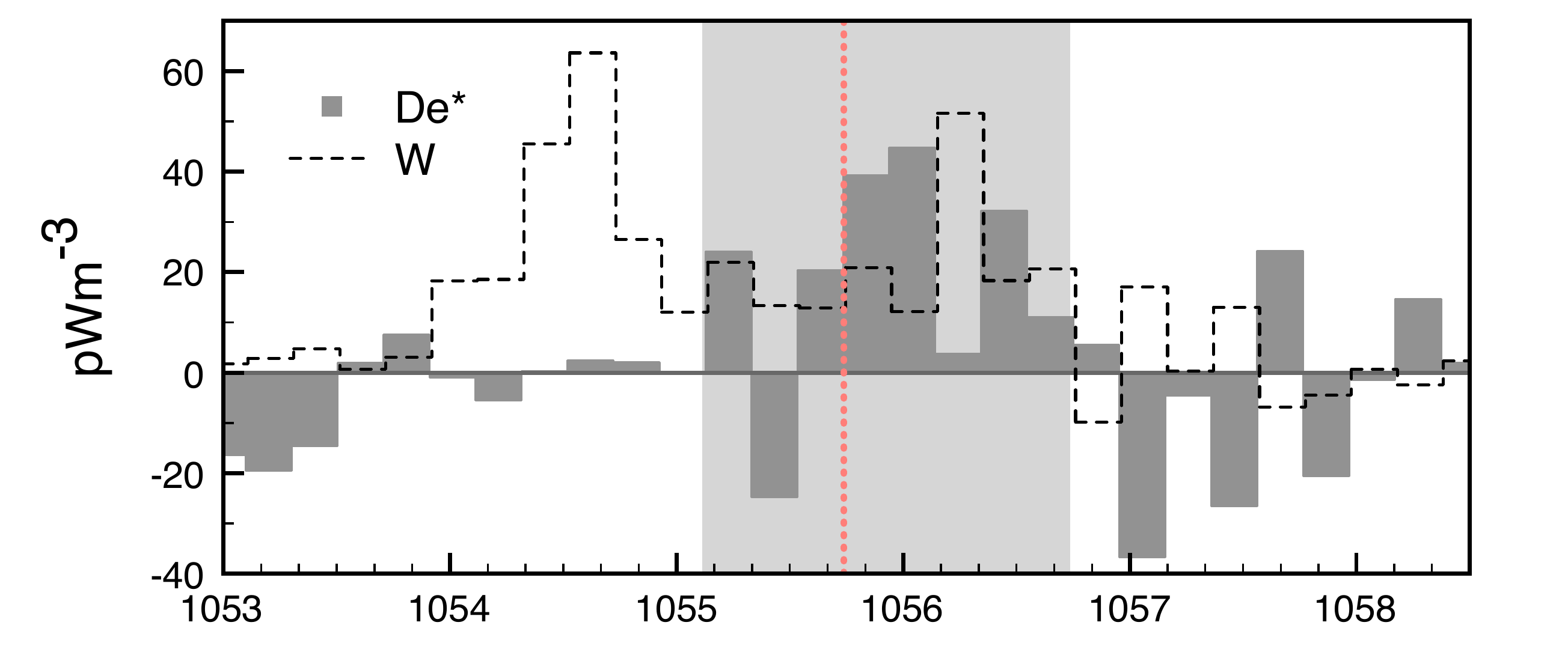}
\caption{(Color online)
The approximate {dissipation} measure $D_e^*$ (gray histogram) and
the work rate by the Lorentz force $W$ (dashed histogram).
The X-point (red dotted line) and the interval (shadow) are
the same as {those} in Figure \ref{fig:geotail}.
}
\end{center}
\label{fig:dissipation}
\end{figure}

Next we introduce several measures to
discuss the reconnection region from Geotail data.
We consider the energy transfer
in the rest frame of the electron's bulk flow.
This measure,
the electron-frame dissipation measure $D_e$,
is formally defined in Eq. (7) in \citet{zeni11c}. 
In a nonrelativistic neutral plasma, 
it can be reduced to
\begin{equation}
\label{eq:EDR2}
D_e = {\bf j} \cdot {\bf E}' = {\bf j} \cdot {\bf E} - W,
\end{equation}
where ${\bf E}'={\bf E}+{\bf v}_e\times{\bf B}$ is the nonideal electric field and
\begin{equation}
\label{eq:W}
W
=
({\bf j} \times{\bf B} ) \cdot {\bf v}_i
=
({\bf j} \times{\bf B} ) \cdot {\bf v}_e
.
\end{equation}
is the work rate per unit volume
by the Lorentz force
to the ion fluid or the electron fluid.
Since this Lorentz work $W$
(/sec /unit volume)
stands for the energy transfer in the ideal MHD,
Equation \ref{eq:EDR2} reads
the total energy transfer
minus the ideal transfer, i.e., 
the {\it nonideal} energy transfer
from fields to plasmas
\citep{birn05,zeni11c}. 
We further assume $E'_z=0$ to obtain the following approximate form,
\begin{equation}
\label{eq:D}
D_e^* = {j}_x {E}'_x + {j}_y {E}'_y.
\end{equation}
The two measures $D_e^*$ and $W$ are shown in Figure \ref{fig:dissipation}.
The current density is directly calculated from the plasma {moments},
${\bf j} = en_i({\bf v}_i-{\bf v}_e)$. 
The dissipation $D_e^*$ (gray histogram) has a peak near the X-point.
Its peak rates are 39 and 45 ${\rm pWm^{-3}}$ 
during 1055:32 -- 1055:44 UT and 1055:44 -- 1055:56 UT (two 12 s intervals), respectively.
The Lorentz work $W$ (dashed histogram) has a peak during 1054:31--1054:43 UT.

\section{PIC simulation}

We validate our observations
with a 2D PIC simulation \citep{zeni11d}.
We employ a Harris-like initial model,
${\bf B}(z)=B_0 \tanh(z/L) {\bf \hat{x}}$ and
$n(z) = n_{0} [0.2 + \cosh^{-2}(z/L)]$, where
$B_0$ and $n_0$ are the reference magnetic field and density.
The current sheet thickness is set to $L=0.5 d_i$,
where $d_i$ is the ion inertial length.
Other simulation parameters are 
$m_i/m_e=100$, $\omega_{pe}/\Omega_{ce}=4$, $T_i/T_e=5$,
and $2400 \times 1600$ cells with $2.2{\times}10^9$ particles. 
The computational domain is
$x, z \in[0,76.8]\times[-19.2, 19.2]$ in units of $d_i$
with periodic ($x$) and reflecting ($z$) boundaries.
Velocities are normalized by the typical Alfv\'{e}n velocity $c_A=B_0/(\mu_0 n_0 m_i)^{1/2}$.

Figure \ref{fig:snapshot} shows
snapshots in the well-developed stage. 
In this case, judging from the sign of $D_e$,
the dissipation region is located around $34.7<x<41.5$ and $-0.38<z<+0.38$.
Its thickness is limited by that of an electron current layer.
The layer thickness is comparable with
the local electron inertial length, $d_e^{\rm loc}\sim 0.31$.
There are also weak dissipative regions at $x=28$ and $x=48$,
where the magnetic field lines suddenly flip.
Surprisingly, both $D_e$ and $D_e^*$ give almost identidal pictures,
as shown in Figures \ref{fig:snapshot}a and \ref{fig:snapshot}b.
There are minor differences near the vertical dissipative regions
but it is hard to recognize them in the figures.

Figure \ref{fig:snapshot}c presents
plasma outflow velocities along the outflow line ($z=0$).
The electron jet substantially overruns
the ion flow \citep{kari07,shay07}
until it reaches a shock-like transition region at $x \approx 28$ and $x \approx 48$.
The electrons are magnetized further downstream.
Note that $v_{ix}$ and $v_{ex}$ are still different in the downstream,
because the ions remain unmagnetized.

Figures \ref{fig:snapshot}d shows
the dissipation measure $D_e$ and the Lorentz work $W$.
The $D_e$ measure is almost
identical to $D_e^*$ along the outflow line,
where $j_z E'_{z}\approx 0$.
Outside the central dissipation region,
$D_e$ exhibits a weak undershoot in the downstream,
indicating energy transfer from plasmas to the electromagnetic fields
in the comoving frame of plasmas.
This is a signature of the fast electron jet region \citep{zeni11d}.
The Lorentz work $W$ is usually positive over the reconnection region,
because the Lorentz force continuously drives plasmas in the outflow directions.
Further downstream,
it has a local peak at $x {\approx} 16.5$ (outside the domain in Fig. 3),
where the reconnected flux ($B_z$) hits the initial current sheet. 

\section{Discussion}

Let us examine the validity of
the approximate measure (Eq. \ref{eq:D}). 
Since EFD does not measure $E_z$ along the spin axis,
we have to reconstruct $E_z$ in Equation \ref{eq:EDR2}.
We consider the following three ways. 
(1) A popular way is to assume ${\bf E}\cdot{\bf B}=0$.
However, 
since there are strong parallel electric field near the dissipation region,
${\bf E}_{\parallel}\ne 0$ \citep{prit01b,wygant05},
and since $B_z$ is usually weak,
this does not work. 
(2) The second is to assume $E_z=0$.
This gives a different picture that emphasizes separatrices. 
Physically, this assumption drops the energy transfer $j_zE_z$ but
leave the ideal part $j_z ({\bf v}_e \times {\bf B})_z$ in Equation \ref{eq:EDR2}.
In other words, the ideal part contaminates the nonideal dissipation measure $D_e$.
Thus we should not assume $E_z=0$.
(3) The third is to set $E_z' = 0$.  This is our choice.
Despite
strong Hall electric field $E_z$ near the dissipation region \citep{shay98,wygant05}, 
the $E'_z\approx 0$ condition is fairly satisfied
outside the electron current layer,
and so $j_zE'_z$ is a minor contributor to $D_e$ along the inflow line
(see Fig. 5a in \citet{zeni11d}).
The ideal condition is also violated $E'_z \ne 0$ along the separatrices
but this does not involve significant energy transfer.
In general, $j_zE'_z$ is negligible due to weak $j_z$.
Thus $D_e^*$ approximates $D_e$ quite well. 
Note that this choice is specific to a magnetotail configuration without a guide field.
In other configurations 
$j_zE'_z$ could be important and Equation \ref{eq:D} may not be useful. 

In Figure \ref{fig:dissipation},
the $D^*_e$ measure has a peak
near the X-point during 1055:44 -- 1056:08 UT.
This is evident for the dissipation region. 
Let us evaluate the dissipation rate.
The reconnection electric field $E_y$ is
fairly uniform over the reconnection region,
except for the magnetic pile-up region near the outflow jet front.
From PIC simulations, we empirically know that
the reconnection rate {remains} constant, $E_y \approx 0.1 c_{A,up}B_{up}$.
Here, $B_{up}$ and $c_{A,up}$ are the magnetic field and the Alfv\'en velocity
in an upstream region, typically a few $d_i$ upstream from the X-point.
It is empirically known that
the half width of the electron current layer is
approximated by the local electron inertial length $d_e^{\rm loc}$
{up to} the case of a realistic mass ratio \citep{prit10b}. 
The current density inside the dissipation region is
$j \sim {B_{d}}/({\mu_0 d_e^{\rm loc}})$,
where $B_{d}$ is the magnetic field at the upstream edge of
the dissipation region.
After the initial current sheet plasma {moves} out,
the plasma density is roughly uniform
over the reconnection region. 
Assuming a uniform density,
one can estimate typical dissipation at the X-point,
\begin{eqnarray}
D_e^*
&\approx& \Big( \frac{B_{d}}{\mu_0 d_e^{\rm loc}} \Big) \Big( 0.1 c_{A,up}B_{up} \Big) 
\nonumber \\
&=& 326
\Big(\frac{B_{d}}{B_{up}}\Big) \Big(\frac{B_{up}}{10 {\rm nT}}\Big)^3
 ~~~{\rm pW~m^{-3}}.
\label{eq:estimate}
\end{eqnarray}
From simulations, we know that
the upstream magnetic field is typically
a half of the asymptotic lobe field : $B_{up}\sim \frac{1}{2} B_{0}$.
In this event, the lobe field is estimated to $B_0=20$ nT,
while $B_x$ occasionally hits 10 nT near the separatricies. 
Thus, it would be reasonable to assume $B_{up}=10$ nT.
Meanwhile, it is very difficult to estimate $B_{d}$,
because the electron-scale dissipation region is embedded
in a thick reconnection layer of
the order of the ion meandering width.
From simulation we obtain $(B_{d}/B_{up})\approx 0.5$.
However, this ratio could be smaller in a real system,
because the electron skin depth with $m_i/m_e=1836$ is 4.3 times smaller than
one with $m_i/m_e=100$.
Using $B_{up}=10$ nT and $(B_{d}/B_{up})<0.5$,
we obtain $D_e^* < 163$ ${\rm pW~m^{-3}}$.
In addition, our data are averaged over a time interval of 12 s. 
Geotail {resolves} the spatial structure rather coarsely.
Using spatially averaged quantities,
their product $D_e^*$ could be underestimated
by a factor of two or three. 
Considering this, the observed rates,
$39$ and $45$ ${\rm pW~m^{-3}}$ (1055:44 -- 1056:08 UT),
are reasonable.

\begin{figure}[thn]
\begin{center}
\includegraphics[width={\columnwidth},clip]{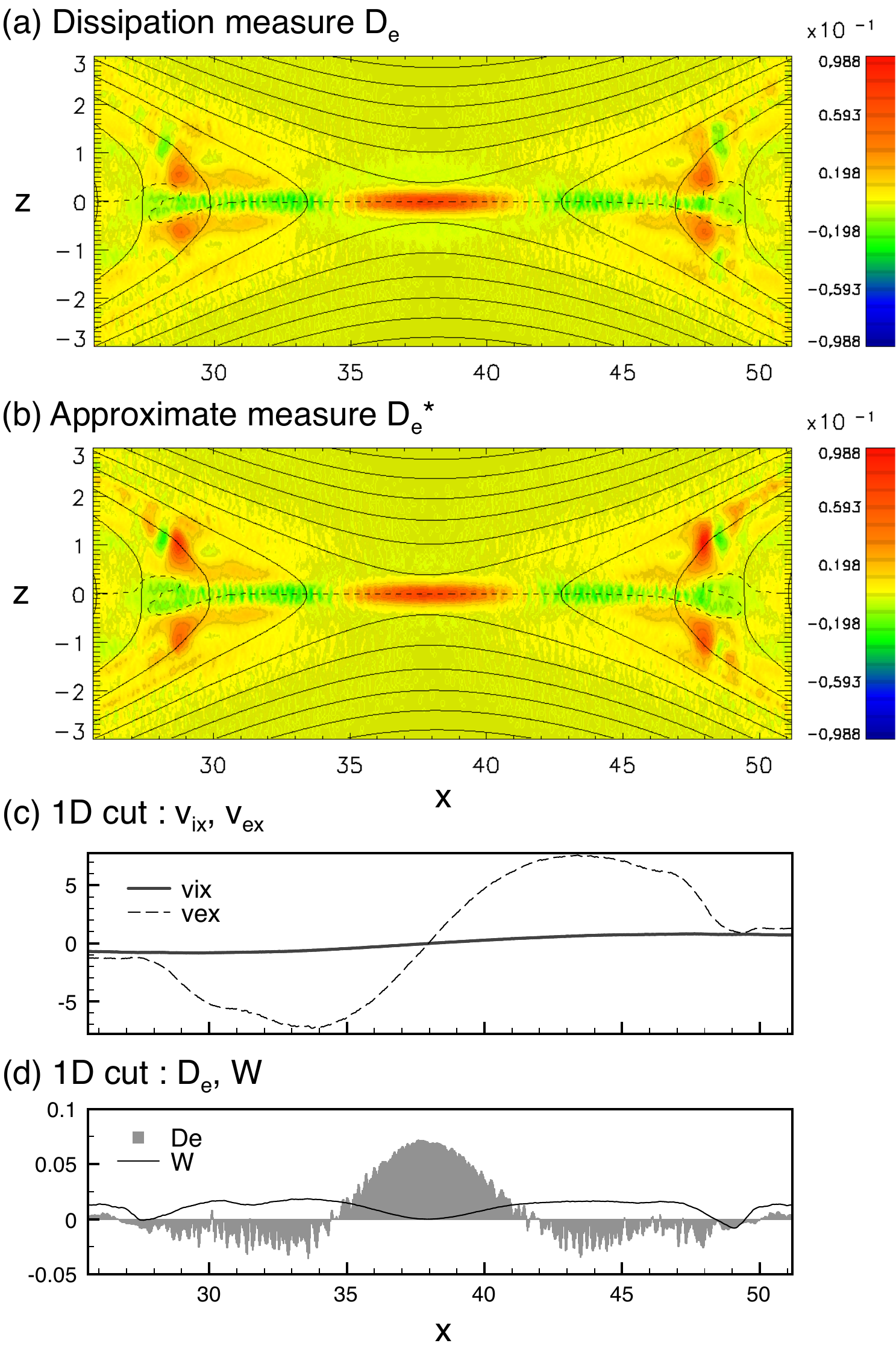}
\caption{(Color online)
\label{fig:snapshot}
(a) The dissipation measure $D_e$ (Eq.~\ref{eq:EDR2}) in unit of $J_0c_AB_0$
with magnetic field lines superposed,
(b) the approximate {dissipation} measure $D_e^*$ (Eq.~\ref{eq:D}) in the same unit,
(c) $v_{ix}, v_{ex}$ at $z=0$, and
(d) the dissipation {measure} $D_e$ and the Lorentz work $W$ at $z=0$.
}
\end{center}
\end{figure}

It is difficult to evaluate unambiguously
the size of the dissipation region
in our single-satellite observation.
\citet{shay07} predicted that
the half length of the $E'_y>0$ region is ${\sim}0.6d_i$.
In many $m_i/m_e=25$ runs [A. Klimas, private communication] and
in our $m_i/m_e=100$ run,
the dissipation region is twice longer than
the $E'_y>0$ region.
Thus the full length of the dissipation region will be $2.4 d_i$.
On the other hand, \citet{zeni11d} reported that
the aspect ratio of the dissipation region is $\sim 0.1$.
Given that the half width of the dissipation region is
$\sim d_e^{\rm loc}$,
the full length will be $\sim 20 d_e^{\rm loc}\sim 0.47 d_i^{\rm loc}$.
In these cases, the full extent of the dissipation region will be
\begin{eqnarray}
2.4 d_i &\sim 1726& \Big( \frac{n_i}{0.1 {\rm cm^{-3}}} \Big)^{-1/2} {\rm km,}
\\
0.47 d_i^{\rm loc} &\sim 1061& \Big( \frac{n_i^{\rm loc}}{0.01 {\rm cm^{-3}}} \Big)^{-1/2} {\rm km}.
\end{eqnarray}
Examining several clues such as the upstream distribution functions,
\citet{nagai11} estimated that the entire structure retreats at ${\sim}100$ km/s.
The above two estimates are comparable with
the spatial scale of one or two sampling intervals (12--24 s), 1200--2400 km.
Geotail detected the dissipation region
by two or three sampling intervals (Fig. \ref{fig:dissipation}),
2400--3600 km or $3$--$5 d_i$ or $1$--$2 d_i^{\rm loc}$.
The length of the dissipation region is on the same order.

The characteristic interval from 1055:07 to 1056:44 UT
(the shadow regions in Figs. \ref{fig:geotail} and \ref{fig:dissipation})
would correspond to
the full extent of the outer electron jet region
in the PIC simulation ($28 < x < 48$ in Fig. \ref{fig:snapshot}).
One can recognize the magnetic field fluctuations
in Figure \ref{fig:geotail}a (indicated by arrows).
It is likely that the fluctuations originate from
the electron jet front region at $x\approx 28$ and $x\approx 48$ (Fig. \ref{fig:snapshot}),
where the fast electron jet hits the outer plasma outflow. 
In observation, $D_e^*$ is negative
before the central dissipation region arrives (Fig. \ref{fig:dissipation}).
This is consistent with
the weak undershoot of $D_e$ in the electron jet region
in the PIC simulation (Fig. \ref{fig:snapshot}). 
We do not recognize a similar undershoot on the other side (1055:44 -- 1056:44 UT).
This is because
the Geotail was near the separatrix,
outside the central electron jet region \citep{nagai11}. 
It is not surprising that
$v_{\perp i}$ still differs from $v_{\perp e}$
outside the interval (Figs. \ref{fig:geotail}c and \ref{fig:geotail}d).
Electrons are magnetized there, but
ions remain unmagnetized downstream of the electron jet fronts. 
Reconnection outflow recovers the ideal MHD approximation
when ions are magnetized further downstream, but
the transition to the MHD region could be ambiguous.

As seen in Figure \ref{fig:dissipation},
the Lorentz work $W$ is usually positive over the reconnection region.
This signature is consistent with the PIC simulation (Section 3).
One can recognize a peak during 1054--1055 UT.
We find that
$B_z$ is compressed (Fig. \ref{fig:geotail}a) in this interval
and that
the plasma density decreased after the $W$-region arrived
(see Fig. 1j in \citet{nagai11}).
This $W$-region appears to be a magnetic pile-up region
behind a tangential discontinuity,
where the compressed $B_z$ pushes the dense plasma sheet outward. 
We do not recognize similar $W$-region on the other side of 1057--1058 UT, 
because the satellite was not in the central neutral plane. 
In fact, we find similar $W$-regions
associated with the pile-up regions in other reconnection events.
The $W$ measure in reconnection deserves
further investigation by theories, simulations, and observations.

To our knowledge,
this is the first identification of
the dissipation region in magnetotail reconnection. 
Earlier observations focused on large-scale signatures of the reconnection site,
such as the flow reversal or Hall fields.
In the 15 May 2003 event, Geotail fortunately resolved
a small-scale signatures of bi-directional electron jets \citep{nagai11} 
and then
the present work finally detected a compact dissipation region between the jets.
Although our resolution is quite limited,
this is the best possible measurement with the present instruments.
In addition, this work is a step forward for reconnection theories.
The dissipation measure theory \citep{zeni11c} was only tested
by 2D PIC simulations with artificial mass ratio.
Here we showed that the theory works in the complex real world.
The theory appears to be practically useful.

\section{Conclusion}

We analyzed Geotail reconnection event on 15 May 2003. 
We applied
the electron-frame dissipation measure $D_e (D^*_e)$
to the reconnection structure. 
Using $D^*_e$, for the first time,
we identified the dissipation region around the possible X-point
in nature.
The dissipation rate ($45 {\rm pWm^{-3}}$) and
the length of the dissipation region (1--2$d_i^{\rm loc}$) are reasonable. 
We introduced the Lorentz work $W$ {as well}.
The two measures are useful for better understanding of the reconnection structure.

Our approach will be applicable to next-generation in situ observations. 
NASA's upcoming Magnetospheric Multiscale (MMS) mission
features high temporal resolution, multi-satellite observation, and
measurement of all three components of the electric field.
We will be able to see clearer pictures of the dissipation region.

\begin{acknowledgments}
S.Z. acknowledges S. Imada, M. Hoshino, Y. Miyashita, M. N. Nishino, and M. H. Saito for useful advices.
\end{acknowledgments}

\end{article}

\end{document}